\documentclass[lettersize,journal]{IEEEtran}

\IEEEoverridecommandlockouts

\usepackage{cite}
\usepackage{amsmath,amssymb,amsfonts}
\usepackage{graphicx}
\usepackage{textcomp}
\usepackage{xcolor}
\usepackage[inline]{enumitem}
\usepackage{subcaption}
\usepackage{easyReview}
\usepackage{array}
\usepackage{float}
\usepackage{multirow}
\usepackage{comment}
\usepackage{mathtools}
\usepackage{booktabs}

\usepackage[hidelinks]{hyperref}

\usepackage{algorithm}
\usepackage{algpseudocode}
\usepackage{bm}

\usepackage{tikz}
\usetikzlibrary{automata,arrows,positioning,calc }

\usepackage{orcidlink}

\def\BibTeX{{\rm B\kern-.05em{\sc i\kern-.025em b}\kern-.08em
    T\kern-.1667em\lower.7ex\hbox{E}\kern-.125emX}}
    
\author{IEEE Publication Technology,~\IEEEmembership{Staff,~IEEE,}
\thanks{This paper was produced by the IEEE Publication Technology Group. They are in Piscataway, NJ.}
\thanks{Manuscript received April 19, 2021; revised August 16, 2021.}}

\usepackage[acronym,nopostdot,  style=super, nonumberlist, toc]{glossaries}
\newacronym[plural=HITLs,firstplural=human-in-the-loop (HITL)]{HITL}{HITL}{human-in-the-loop}
\newacronym[plural=CPSs,firstplural=cyber-physical systems (CPS)]{CPS}{CPS}{cyber-physical system}
\newacronym[]{TSN}{TSN}{time sensitive networking}
\newacronym[plural=CPSs]{UE}{UE}{user equipment}
\newacronym{HARQ}{HARQ}{Hybrid-ARQ}
\newacronym{MCS}{MCS}{modulation coding scheme}
\newacronym{MLP}{MLP}{multi-layer perceptron}
\newacronym{SDR}{SDR}{software-defined radio}
\newacronym{OAI}{OAI}{OpenAirInterface}
\newacronym{SINR}{SINR}{signal to interference and noise ratio}
\newacronym{LSTM}{LSTM}{long short-term memory networks}
\newacronym{RNNs}{RNNs}{recurrent neural networks}
\newacronym{MAE}{MAE}{mean absolute error}
\newacronym{NLL}{NLL}{negative loglikelihood}
\newacronym{RNN}{RNN}{recurrent neural network}
\newacronym{CDF}{CDF}{cumulative distribution function}
\newacronym{RTT}{RTT}{round-trip time}
\newacronym{PTP}{PTP}{precision time protocol}
\newacronym{CDE}{CDE}{conditional density estimation}
\newacronym{GMM}{GMM}{Gaussian mixture model}
\newacronym{GMEVM}{GMEVM}{Gaussian mixture extreme value model}
\newacronym{TDD}{TDD}{time division duplex}
\newacronym[plural=PRBs,firstplural=physical resource blocks (PRB)]{PRB}{PRB}{physical resource block}
\newacronym[plural=RANs,firstplural=radio access networks (RAN)]{RAN}{RAN}{radio access network}
\newacronym{COTS}{COTS}{commercial off-the-shelf}
\newacronym[plural=MDNs,firstplural=mixture density networks (MDN)]{MDN}{MDN}{mixture density network}
\newacronym[plural=MLEs,firstplural=maximum likelihood estimation (MLE)]{MLE}{MLE}{maximum likelihood estimation}
\newacronym[plural=KL-divergences,firstplural=Kullback-Leibler divergences (KL-divergence)]{KL-divergence}{KL-divergence}{Kullback-Leibler divergence}
\newacronym[]{EVT}{EVT}{extreme value theory}
\newacronym[plural=GPDs,firstplural=generalized Pareto distributions (GPD)]{GPD}{GPD}{generalized Pareto distribution}
\newacronym[plural=PDFs,firstplural=probability density functions (PDF)]{PDF}{PDF}{probability density function}
\newacronym{ARIMA}{ARIMA}{autoregressive integrated moving average}
\newacronym{V2X}{V2X}{vehicle-to-everything}
\newacronym{QoS}{QoS}{quality of service}
\newacronym{mmWave}{mmWave}{millimeter wave}
\newacronym{IoT}{IoT}{internet of things}
\newacronym{CPU}{CPU}{central processing unit}
\newacronym{URLLC}{URLLC}{ultra reliable low latency communications}
\newacronym{RLC}{RLC}{radio link control}
\newacronym{NACK}{NACK}{negative acknowledge}

\usepackage{amsmath,amsfonts} 
\usepackage{graphicx}

\begin{document}

\title{Probabilistic Delay Forecasting in 5G Using Recurrent and Attention-Based Architectures}

\author{
Samie Mostafavi \orcidlink{0000-0001-9316-0414}
,~\IEEEmembership{Member,~IEEE,}
Gourav Prateek Sharma \orcidlink{}
,~\IEEEmembership{Member,~IEEE,}
Ahmad Traboulsi \orcidlink{0009-0009-7802-1964}
,~\IEEEmembership{Member,~IEEE,}
James Gross \orcidlink{0000-0001-6682-6559}
 ,~\IEEEmembership{Senior Member,~IEEE}
\thanks{This work was supported by the European Commission through the H2020 project DETERMINISTIC6G (Grant Agreement no. 101096504).}
\thanks{ Samie Mostafavi, Gourav Prateek Sharma, Ahmad Traboulsi and James Gross are with the KTH Royal Institute of Technology, Stockholm, Sweden (e-mails: ssmos@kth.se; gpsharma@kth.se; ahmadtr@kth.se; jamesgr@kth.se)}
}

\maketitle

\thispagestyle{plain}
\pagestyle{plain}

\begin{abstract}

With the emergence of new application areas—such as cyber-physical systems and human-in-the-loop applications—ensuring a specific level of end-to-end network latency with high reliability (e.g., 99.9\%) is becoming increasingly critical. To align wireless links with these reliability requirements, it is essential to analyze and control network latency in terms of its full probability distribution. However, in a wireless link, the distribution may vary over time, making this task particularly challenging.
We propose predicting the latency distribution using state-of-the-art data-driven techniques that leverage historical network information. Our approach tokenizes network state information and processes it using temporal deep-learning architectures—namely LSTM and Transformer models—to capture both short- and long-term delay dependencies. These models output parameters for a chosen parametric density via a mixture density network with Gaussian mixtures, yielding multi-step probabilistic forecasts of future delays.
To validate our proposed approach, we implemented and tested these methods using a time-synchronized, SDR-based OpenAirInterface 5G testbed to collect and preprocess network-delay data.  Our experiments show that the Transformer model achieves lower negative log-likelihood and mean absolute error than both LSTM and feed-forward baselines in challenging scenarios, while also providing insights into model complexity and training/inference overhead. This framework enables more informed decision-making for adaptive scheduling and resource allocation, paving the way toward enhanced QoS in evolving 5G and 6G networks.

\end{abstract}


\section{Introduction}
\label{sec:introduction}
\IEEEPARstart{T}{raditionally}, communication networks have been built around a best‐effort model, offering no formal performance guarantees. However, emerging applications such as cyber-physical systems and human-in-the-loop scenarios demand predictable, low-latency connectivity to ensure safe and correct operations. In these systems, information is exchanged between components like physical sensors, actuators, and computing devices to support applications ranging from industrial automation to virtual reality. Typically, for such applications, end-to-end latency which is the time it takes for a packet to travel from a sensor to a controller and back to an actuator, must be maintained within 1 to 50 ms and delivered with exceptionally high reliability (e.g., $\geq$99.99\%). In such contexts, even minor delays can compromise system responsiveness and safety.
The emergence of \gls{TSN} standards under IEEE 802.1, which define traffic-shaping mechanisms to achieve bounded latency in switched Ethernet networks, reflects the growing demand for predictable, low-latency solutions across a range of advanced applications \cite{nasrallah2018ultra}.

Recent enhancements like Ultra-Reliable Low-Latency Communications (URLLC) have strived to narrow the gap between application demands and actual wireless performance \cite{ansari2022performance}. However, meeting precise latency bounds in a probabilistic sense remains an open problem, as even small deviations from target deadlines can severely impact time-critical tasks.
To address this challenge, an essential first step is to characterize or predict network latency in a probabilistic manner. By accurately modeling and forecasting the full latency distribution, system designers can then explore advanced resource allocation and scheduling mechanisms that more reliably meet strict latency requirements, ultimately enabling robust performance guarantees in future 5G and 6G networks \cite{mostafaviaqm2023, egger2025}.

\begin{figure}[ht]
    \centering
    \includegraphics[width=0.48\textwidth, trim={1cm 0.4cm 0.8cm 0.3cm}, clip]{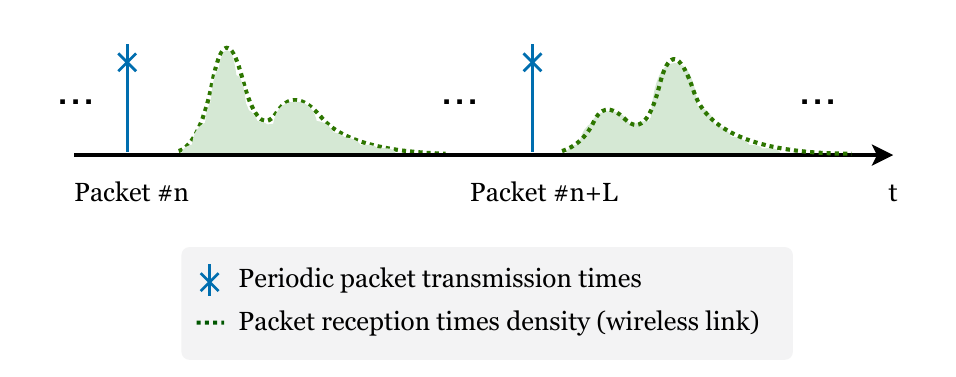}
    \caption{Illustration of packet transmission cycles over a wireless link, showing a broad reception time profile caused by the stochastic nature of wireless delays, which may change over time from cycle $n$ to $n+L$}
    \label{fig:intro}
\end{figure}

However, probabilistic latency characterization in wireless systems is complicated by a wide, evolving distribution driven by factors such as interference, channel fading, and mobility, as depicted in Figure~\ref{fig:intro}. These conditions can introduce temporal dependencies, demanding advanced modeling methods to capture the dynamic nature of delay—an area that remains underexplored in communications research. Meanwhile, the machine learning community has produced a new wave of time-series forecasting tools (e.g., RNNs, LSTMs, and Transformers), which enable large-scale, probabilistic predictions \cite{hochreiter97, vaswani2017}. Integrating these techniques allows us to forecast the full latency distribution over multiple future time steps, thereby capturing both short- and long-term dependencies. Additionally, because end-to-end delays hinge on a diverse set of network parameters, selecting and processing the most pertinent features is non-trivial. Tokenization strategies derived from large language models (LLMs) have demonstrated remarkable potential for compactly representing high-dimensional, heterogeneous data \cite{devlin2019bertpretrainingdeepbidirectional}. Consequently, we view them as a compelling approach for accurately modeling and forecasting latency dynamics.

Numerous studies have employed data-driven methods for delay prediction in wireless communication networks, as they provide significant flexibility—especially when deep learning is used to model the complex behaviors inherent in such systems. 
Among them, simulation-based studies play a vital role in evaluating data-driven delay prediction techniques. For example, Moreira et al. \cite{moreiraQoSPredictabilityV2X2020} compared different methods—including \glspl{MLP}, random forests, and \gls{ARIMA}—for point delay estimates in a simulated 5G vehicular scenario. Similarly, Barmpounakis et al. \cite{barmpounakisLSTMbasedQoSPrediction2021} utilized NS3 simulations for packet latency prediction in a \gls{V2X} context by applying \gls{LSTM} networks, while Dang et al. \cite{dangTimeDelayPrediction2023} applied bidirectional \glspl{LSTM} to estimate point delays and introduced optimization policies to improve performance in a simulated 5G environment.
Although simulation environments provide a controlled setting to develop and test these methods, their reliance on simplifying assumptions can limit their practical validity. Real-world data is indispensable because it captures the full complexity and unpredictability of wireless networks, ensuring that predictive models remain robust under operational conditions.

A number of studies have been conducted on real wireless network data, where many works focus on point estimates or the estimation of specific quantiles of delay \cite{khatouniMachineLearningApplication2019, raoPredictionExposureDelays2022, raoGeneralizableOneWayDelay2024, kulzerLatencyPredictionDeep2021}. Notably, Rao et al. \cite{raoGeneralizableOneWayDelay2024} concentrated on one-way delay prediction using domain adaptation techniques to improve generalizability across unseen user equipment, training neural networks on a real 5G \gls{mmWave} testbed. In another study, Palaios et al. \cite{palaiosMachineLearningQoS2023a} collected highway data from connected vehicles in a test LTE network to measure throughput and latency, employing models such as decision trees and \glspl{MLP} for point delay prediction. Other works have addressed temporal dependencies in delay; for instance, in \cite{LatencyPredictionDelaysensitive}, an \gls{LSTM}-integrated framework was proposed for point delay estimation in both fixed and mobile scenarios within LTE networks. More recently, attention-based autoregressive models—such as Transformers—have been used for \gls{QoS} prediction, with studies like \cite{yimengPredictionMethod5G2022} and \cite{azmin2022bandwidth} demonstrating the superiority of Transformers over traditional approaches (e.g., \gls{LSTM} and \gls{ARIMA}) for predicting wireless channel characteristics on real datasets.
Despite these advances, point estimates are often insufficient for many delay-sensitive applications that require a complete probabilistic characterization of delay. 

Methods capable of capturing the full stochastic nature of delay provide richer information that is more useful for ensuring robust performance in time-sensitive applications. Some works have addressed probabilistic delay prediction on real wireless data using non-conditional approaches. For example, Volos et al. \cite{volosLatencyModelingMobile2018} employed mixture models on real LTE delay measurements, demonstrating that mixtures of lognormals and extreme value theory models (such as the \gls{GPD}) can more effectively model rare latency events. Similarly, Fadhil et al. \cite{fadhilEstimation5GCore2022} investigated the latency profile in 5G networks using Gaussian mixture models, examining how increasing the number of mixture centers influences fit accuracy.

Further refinement of probabilistic delay prediction is achieved when side information is incorporated and the problem becomes conditional density estimation. Flinta et al. \cite{flinta_predicting_2020} used random forests and a histogram-based approach to condition delay distribution predictions on contextual variables such as position, time, radio channel, and signal power in \gls{IoT} networks. In a related study targeting cloud environments, Samani et al. \cite{samaniConditionalDensityEstimation2021a} combined deep neural networks with Gaussian mixture models in the framework of \glspl{MDN} to predict service metrics from conditions such as \gls{CPU} utilization.
Moreover, in our previous work \cite{mostafaviDataDrivenLatencyProbability2023c}, we applied mixture models to real 5G delay measurements using MCS indices as conditioning variables, demonstrating that mixtures of Gaussians and extreme value theory models (such as the \gls{GPD}) are particularly effective in modeling rare latency events.

Temporal probabilistic prediction on real-world data is still a relatively underexplored area. Skocaj et al. \cite{skocajDatadrivenPredictiveLatency2023a} leveraged \glspl{RNN} and \glspl{LSTM} on mobile network operator data to derive conditional delay probability density functions that account for dependencies on network and traffic conditions. Nonetheless, their method is confined to single-step predictions, and its sensitivity and efficiency were not rigorously assessed, positioning it primarily as a proof of concept.

A significant gap in the current literature (and one that this paper addresses) is the development of probabilistic delay forecasting, which is crucial for proactive network operations. While many approaches predict delay for only a single future time step, the models that forecast temporal dependencies across multiple time steps can extend the utility of predictions, enabling network operators to plan for a longer future horizon. 
Another observation from the literature is the prevalent reliance on manual feature selection for data encoding and embedding. Such approaches can impede scalability.
In summary, while data-driven delay prediction frameworks have advanced considerably—from simulation-based evaluations to models trained on real-world data—there remains a need for methods that offer multistep probabilistic forecasts and scalable data encoding. These are precisely the challenges that this paper seeks to address.

In this work, we advocate for a deep temporal neural network strategy that leverages tokenization of contextual packet information for probabilistic latency prediction. 
Our primary contributions in this work are summarized as follows:
\begin{itemize} 

\item \textbf{Temporal Delay Dependency Modeling:} We introduce a novel probabilistic delay prediction framework that captures temporal dependencies in 5G networks. By embedding historical packet delay and contextual network state information into fixed-dimensional tokens and processing these with advanced deep temporal models (e.g., LSTM and Transformer architectures), our approach effectively learns the dynamic evolution of network conditions.

\item \textbf{Multi-Horizon Prediction Capability:} We propose and study a multi-horizon prediction methodology that forecasts the entire delay distribution for several future packets simultaneously. This multi-step prediction, which is largely absent in existing literature, provides a comprehensive view of future network behavior.

\item \textbf{Systematic Delay Context Encoding:} We employ a tokenization approach that systematically encodes a diverse range of network context information—including instantaneous packet data and historical trends—into a fixed-dimensional representation. This encoding facilitates effective representation learning, allowing the temporal models to automatically discover and leverage the most relevant features for accurate delay distribution prediction, without manual feature selection.

\item \textbf{Real-world Implementation, Evaluation, and Learning Efficiency Analysis:} We deploy and rigorously evaluate several delay predictor variants on a cutting-edge SDR-based 5G testbed. In addition to measuring prediction accuracy under realistic network scenarios, we also analyze learning efficiency in depth. Specifically, we examine training time and data requirements, addressing the computational demands common in time series forecasting.

\item \textbf{Quantitative Performance Gains:} Through extensive evaluations on real 5G testbed data, we demonstrate that our multi-step Transformer approach consistently outperforms both single-step and standard recurrent baselines in terms of negative log-likelihood and mean absolute error, highlighting its ability to accurately capture the evolving latency distribution over longer horizons and under varied channel conditions.

\end{itemize}

The rest of the paper is organized as follows: Section 2 presents the system model and the probabilistic prediction framework, Section 3 presents our machine learning-based approaches, Section 4 elaborates on the methodology of this research, while Section 5 contains the numerical evaluations and assessment of results.
Finally, Section 6 concludes this work.
\section{System Description and Problem Formulation}
\label{sec:system_model}

\begin{figure}[ht]
    \centering
    \includegraphics[width=0.48\textwidth, trim={0.7cm 0.4cm 0.7cm 0.4cm}, clip]{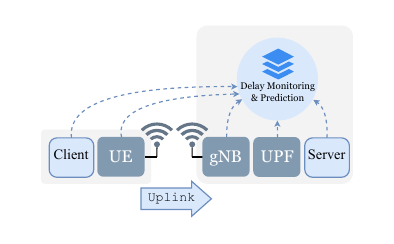}
    \caption{A 5G uplink system model illustrating the end-to-end data path and the real-time delay monitoring \& prediction module gathering relevant information to estimate the one-way delay.}
    \label{fig:systemmodel}
\end{figure}

As shown in Figure~\ref{fig:systemmodel}, we examine a 5G uplink scenario where one-way delay (OWD) is critical for applications such as robot sensor data transmissions. These applications typically involve the periodic exchange of fixed-size packets under stringent delivery constraints~\cite{craciunas2016scheduling}. Consequently, we assume that packets of size \(B_s\) arrive at regular intervals \(T_s\). Given the inherent randomness of wireless networks, we model the end-to-end latency of each packet \(n\) as a random variable \(Y_n \in \mathbb{R}_{+}\). Importantly, these latencies can exhibit serial correlation, capturing the stochastic and time-varying nature of wireless channels.
 
As a packet traverses the link, it is exposed to a range of conditions—such as channel quality, scheduling delays, and buffering—that directly or indirectly affect its end-to-end delay.
We consider a real-time monitoring system that gathers these information to be used for training or prediction in the delay predictor.
This data for each packet \( n \) is an \( M \)-dimensional random vector \( X_n \) (with realization \( \mathbf{x}_n \)), which we refer to as the \emph{packet delay context vector}.
Among the possible elements of $X_n$, two principal sources of stochastic delay in wireless communications are particularly significant, as detailed below for subsequent analysis:
\begin{itemize} 
    \item \textbf{Channel-Induced Delays}: Packets are segmented into code blocks and transmitted using a specific \gls{MCS}. Decoding failures at the receiver, caused by channel noise or fading, trigger retransmissions via the \gls{HARQ} protocol, introducing stochastic delays until the entire packet is successfully decoded. Thus, in addition to the \gls{MCS} index, the total number of \gls{HARQ} retransmissions per packet is a critical factor in the overall delay. Moreover, if a block fails to decode after typically four \gls{HARQ} attempts, it is deemed unsuccessful and the \gls{RLC} layer sends a \gls{NACK} to restart the process at a higher layer. This reinitialization, involving control messages in the RLC layer, usually results in significantly larger and more variable delays depending on channel conditions. Consequently, the number of RLC retransmissions during packet transmission is also an important factor to capture and analyze. 
    \item \textbf{Scheduling-Induced Delays}: Schedulers might defer transmissions or allocate insufficient resources, potentially causing packet segmentation and increased latency. Such delays can result from resource contention or misalignments in the \gls{TDD} slot allocations of the cellular system. For instance, packets arriving during an incompatible \gls{TDD} phase (e.g., an uplink packet arriving during a downlink slot) may experience queuing delays, while resource constraints may force the scheduler to postpone transmissions. In this case, we consider the timeslot in which the packet arrives on the 5G system as another key factor when predicting delay.
\end{itemize}
While existing literature offers insights into which factors are most crucial, our model is agnostic to the specific features used. In our work, we primarily consider metrics such as \gls{MCS} index, the number of \gls{HARQ} and \gls{RLC} retransmissions, packet size and periodicity, and the 5G slot in which the packet arrive.
Additionally, the same framework and methodology can be readily extended to downlink or multi-hop scenarios, provided the necessary data is available.

Beyond these stochastic elements, packet delays also exhibit serial correlation.  Significant delays experienced by packet \(n\) can propagate to subsequent packets (\(n+1, n+2, \dots\)) due to buffering mechanisms within the 5G network. This effect becomes more pronounced with smaller packet inter-arrival times or increased average packet delays. Furthermore, persistent conditions like poor channel quality or sustained network congestion create temporal correlations in the delay profile.  These autocorrelations manifest as periods of prolonged congestion, recurring unfavorable arrival times relative to the \gls{TDD} structure, or extended periods of poor channel conditions, leading to correlated delay patterns.

During a sequence of packet transmissions, upon the arrival of packet \(n\), our objective is to predict not only its delay distribution but also the delay distributions of the subsequent packets \(n+1, \dots, n+L-1\) using historical observations from the \(H\) previous packet transmissions and their delay contexts (including that of packet \(n\)), i.e., packets \(n, n-1, \dots, n-H+1\). Hence, we consider the previously observed \(H\) packet delay context vectors up to the arrival of packet \(n\) to estimate the \(L\) conditional probability distributions:
\begin{equation}
    P\bigl(Y_{n+l} \mid X_{n-H+1:n} = \mathbf{x}_{n-H+1:n}\bigr),
\end{equation}
for \(l=0,1,\dots, L-1\), where the notation \(X_{n-H+1:n}\) denotes the collection of previous \(H\) packet delay context vectors from \(n-H+1\) to \(n\).

In the following section, we detail our proposed approach, describing how we model these dependencies and predict the packet delay distributions.

\section{Approach} \label{sec:approach}

\begin{figure}[ht]
    \centering
    \includegraphics[width=0.48\textwidth, trim={2.3cm 0.4cm 1.2cm 0.3cm}, clip]{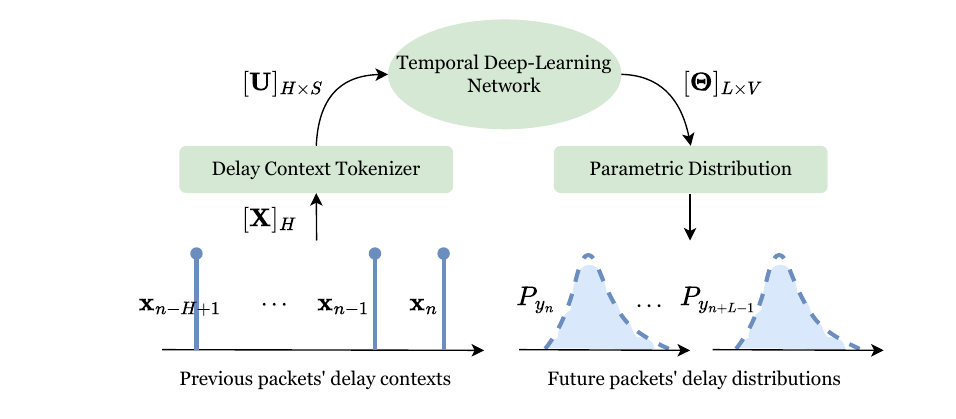}
    \caption{Temporal-based learning and prediction approach overview}
    \label{fig:approach}
\end{figure}

In this section, we introduce our methodology for predicting the end-to-end delay distribution of future packets with temporal and contextual dependencies. 
Given the impracticality of deriving closed-form solutions, we adopt a data-driven approach. 
We leverage observed packet delay histories leading up to a packet’s arrival to forecast its delay and those of future arrivals (up to $L$ packets ahead) to predict time-dependent probabilistic behavior of packet delay.

The proposed approach consists of three main steps: (i) encoding historical observations into tokens, (ii) leveraging a temporal deep-learning architecture (e.g. \gls{LSTM} or Transformer) to capture time-dependent patterns, and (iii) generating the parameters of a chosen parametric density function to obtain the delay distribution. Below, we describe each component in detail.

\subsection{Tokenization of Historical Observations}

To incorporate historical data into a deep-learning model, we first map each packet delay context $\mathbf{x}_n$ to a fixed-dimensional token $\mathbf{u}_n \in \mathbb{R}^{S}$ using a learnable embedding function $\phi$. This function concatenates individual feature embeddings (e.g., packet size, periodicity, \gls{MCS} index, retransmissions, queue length) and processes them through a \gls{MLP} to produce the $S$-dimensional representation $\mathbf{u}_n$. 
By training $\phi$ jointly with the temporal model, we automatically extract relevant features and eliminate extensive manual engineering. Next, we gather the most recent $H$ tokens $\{\mathbf{u}_{n-H+1}, \ldots, \mathbf{u}_n\}$ and stack them into $\mathbf{U} \in \mathbb{R}^{H \times S}$, which serves as input to the temporal architecture.
This tensor captures the immediate past network states and delays, enabling the model to infer evolving delay levels from channel conditions and scheduling dynamics.

\subsection{Temporal Deep-Learning Architecture}

Once the embedding tensor \(\mathbf{U} \in \mathbb{R}^{H \times S}\) is formed by stacking \(H\) tokenized observations, the next step is to capture temporal dependencies and generate predictions for future packet delays. In what follows, we describe two commonly used deep-learning frameworks for this task: (i) a recurrent architecture (e.g. \gls{LSTM}) and (ii) an attention-based architecture (e.g. Transformer). Our goal in this part of the approach is to produce a tensor \(\boldsymbol{\Theta} \in \mathbb{R}^{L \times V}\), where \(L\) is the number of future packets for which we want to predict the delay distribution, and \(V\) is the dimensionality of the parametric density function's parameters (e.g. mean, variance, and weights for a \gls{GMM}).
The result is going to be $\boldsymbol{\Theta} = \{ \boldsymbol{\theta}_{n+i} \}_{i=0}^{L-1} \in \mathbb{R}^{L \times V}$, where each \(\boldsymbol{\theta}_{n+l}\) is a vector of size $V$ and fully characterizes the predicted distribution for the delay of packet \(n+l\) for $l \in \{0:L-1\}$.

Next, we describe the two proposed approaches for converting a sequence of delay tokens into a sequence of distribution parameters: one that uses a recurrent architecture (e.g., LSTM) and another that employs an attention-based architecture (e.g., Transformer).
While both LSTMs and Transformers have proven effective in capturing temporal dependencies, they possess distinct characteristics. LSTMs, renowned for their ability to handle sequential data, excel at modeling long-term dependencies through their memory cells. However, they can be computationally expensive, especially for long sequences. Transformers, on the other hand, leverage self-attention mechanisms to weigh the importance of different input tokens, enabling them to capture complex relationships between distant elements. While Transformers often outperform LSTMs in tasks requiring long-range dependencies, they can be less efficient for shorter sequences. Given the varying nature of packet delay patterns, which can exhibit both short-term fluctuations and long-term trends, we believe that a comparative analysis of both architectures is essential to identify the most suitable approach for our problem \cite{hochreiter97, vaswani2017}.

\subsubsection*{Recurrent Neural Architecture}

\begin{figure}[ht]
    \centering
    \includegraphics[width=0.48\textwidth, trim={2.5cm 0.4cm 2.5cm 0.4cm}, clip]{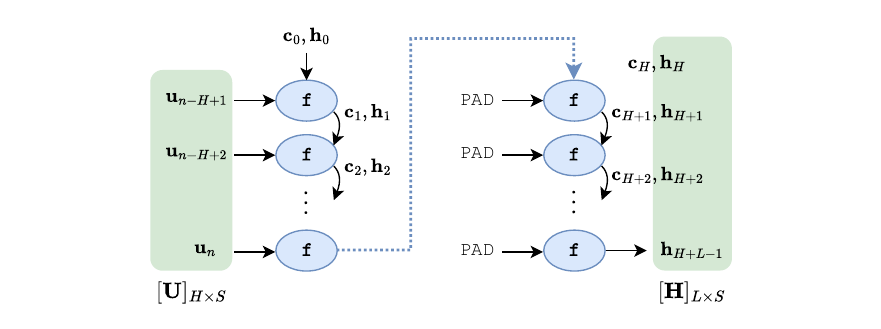}
    \caption{Recurrent-based neural prediction}
    \label{fig:approach_rnn}
\end{figure}

A recurrent neural network, such as an \gls{LSTM}, processes the sequence \(\mathbf{U}\) \emph{sequentially}. As shown in Figure \ref{fig:approach_rnn}, at each time step \(i\) (\(i = 1, \dots, H\)), the \gls{LSTM} cell updates its hidden and cell states according to:
\begin{equation}
    (\mathbf{h}_i, \mathbf{c}_i) = \mathtt{f}\bigl(\mathbf{u}_{n-H+i}, \mathbf{h}_{i-1}, \mathbf{c}_{i-1}\bigr).
\end{equation}
Here, \(\mathbf{h}_i\) is the hidden state used as the output at each time step, and \(\mathbf{c}_i\) is the cell state, which serves as a long-term memory capable of retaining information over multiple steps. Both of these states evolve under trainable weights associated with the input, forget, and output gates. The initial states \(\mathbf{h}_0\) and \(\mathbf{c}_0\) may be set to zero or learned. After processing all \(H\) tokens, the final hidden state \(\mathbf{h}_H\) acts as a summarized representation of the historical context.

In practice, multiple \(\mathrm{LSTM}\) or \(\mathrm{RNN}\) layers are often \emph{stacked} to increase the model’s capacity for learning complex patterns. For example, in a two-layer LSTM, the output of the first layer at each step \(i\) (i.e., \(\mathbf{h}_i^{(1)}\)) serves as the input to the second layer:
\begin{equation}
    (\mathbf{h}_i^{(2)}, \mathbf{c}_i^{(2)}) = \mathtt{f}^{(2)}\bigl(\mathbf{h}_i^{(1)}, \mathbf{h}_{i-1}^{(2)}, \mathbf{c}_{i-1}^{(2)}\bigr),
\end{equation}
where the superscript \({(j)}\) denotes the layer index \(j \in \{1, 2\}\). This process can be repeated for additional layers, typically up to two or three in many applications. 
After the sequence of length \(H\) is processed in each layer, the last hidden state of the topmost layer \(\mathbf{h}_H^{(N_{\mathrm{rec}})}\) (where \(N_{\mathrm{rec}}\) is the total number of stacked layers) acts as the final historical context embedding.

To predict delays for the next \(L\) packets, we employ the autoregressive decoding scheme to produce \(\boldsymbol{\Theta} \in \mathbb{R}^{L \times V}\), where each row corresponds to one future time step's parametric density parameters.
We use the final hidden states \(\mathbf{h}_H,\mathbf{c}_H\) as the initial states for subsequent unrolling of the recurrent cell for \(L-1\) steps. At each step \(l \in \{1,\dots,L-1\}\), the cell generates a hidden state \(\mathbf{h}_{H+l}\), which is passed through a fully connected layer to produce \(\boldsymbol{\theta}_{n+l} \in \mathbb{R}^V\). These parameters correspond to the predicted density function for packet \(n+l\).
As illustrated in Figure \ref{fig:approach_rnn}, we use padding tokens on the inputs for the generation phase as the future steps are unknown.

\subsubsection*{Transformer Architecture}

\begin{figure}[ht]
    \centering
    \includegraphics[width=0.48\textwidth, trim={1.0cm 0.4cm 2.8cm 0.3cm}, clip]{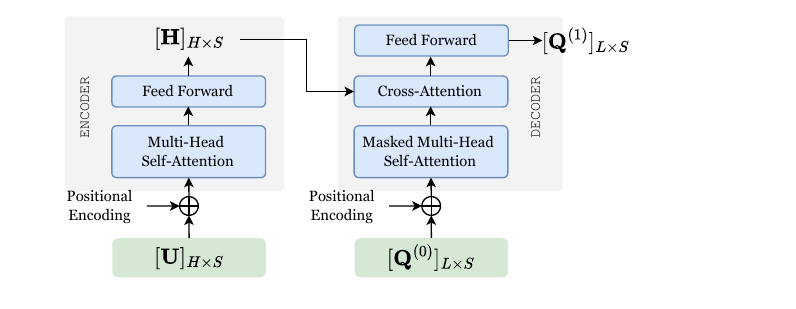}
    \caption{Transformer-based neural prediction}
    \label{fig:approach_tf}
\end{figure}

Transformers, in contrast to recurrent-based models, rely on self-attention mechanisms to capture dependencies across the entire input sequence. The key building block is the \emph{multi-head self-attention} module, which weighs the relevance of each input token against every other token to compute a contextual representation.
A Transformer encoder layer comprises:
\begin{enumerate}
    \item \emph{Multi-head self-attention}: Computes attention weights between all pairs of tokens \(\mathbf{u}_{n-i}\) and \(\mathbf{u}_{n-j}\) for \(i,j \in \{0,\dots,H-1\}\).
    \item \emph{Feed-forward network}: A position-wise fully connected network applied to each token embedding after self-attention.
    Residual connections and layer normalization in this layer facilitates stable training by preserving gradient flow and normalizing activations.
\end{enumerate}
As shown in Figure \ref{fig:approach_tf}, prior to applying self-attention, positional encodings are added to the embedded tokens \(\mathbf{U}\) to inject information about the ordering in the sequence. The Transformer encoder processes each row (token) in parallel: \(\mathbf{H}^{(1)} = \mathtt{g}^{\text{enc},(1)}(\mathbf{U})\), where $\mathtt{g}^{\text{enc},(j)} : \mathbb{R}^{H \times S} \rightarrow \mathbb{R}^{H \times S} $ is a function representing the encoder layer $j$ and  \(\mathbf{H}^{(1)} \in \mathbb{R}^{H \times S}\) is the hidden representation after the first encoder layer. Stacking multiple encoder layers refines the representations further:
\begin{equation}
    \mathbf{H}^{(j+1)} 
    = 
    \mathtt{g}^{\text{enc},(j+1)}(\mathbf{H}^{(j)}),
    \quad
    j = 1, \dots, N_{\mathrm{enc}}-1,
\end{equation}
where \(N_{\mathrm{enc}}\) is the number of layers in the encoder.

To predict the next \(L\) time steps, our framework adopts an encoder-decoder architecture rather than reusing the same recurrent cell. The encoder first converts the historical sequence into a contextual representation, denoted by \(\mathbf{H}^{(N_{\mathrm{enc}})} \in \mathbb{R}^{H \times S}\). The decoder then leverages this representation to generate the forecast parameters \(\mathbf{\Theta} \in \mathbb{R}^{L \times V}\) for the upcoming time steps.
As shown in Figure \ref{fig:approach_tf}, Transformer decoders rely on an architecture that requires a dedicated embedding space and work with a separate \emph{target} sequence. We denote this target sequence by \(\mathbf{Q} \in \mathbb{R}^{L \times S}\), where \(L\) is the number of future time steps and \(S\) is the embedding dimension. Each row of \(\mathbf{Q}\) represents a learnable embedding (analogous to input tokens) and serves as a placeholder for the future outputs. The motivation behind this design is that the decoder learns to generate the next token in the target space conditioned on its own previous predictions, thereby capturing temporal relationships effectively within this dedicated embedding space.

In standard Transformer implementations, decoders typically operate in an autoregressive mode. At inference time, the decoder generates tokens one-by-one, using its own previous outputs as inputs for subsequent steps. This sequential approach is beneficial in capturing complex, long-range dependencies within the output space, but it is inherently slower because each token must wait for the previous one to be generated.

Alternatively, the decoder can be run in a parallel (non-autoregressive) mode, where all output tokens are generated simultaneously. In this approach, placeholders (or appropriate padding) are provided for positions that would otherwise depend on earlier predictions. The primary advantage of parallel decoding is its speed, as it eliminates the need to wait for sequential outputs. In our application---modeling periodic packet transmissions---the temporal correlations in the output space are relatively simple, so we hypothesize that a parallel decoding approach will suffice. (This point will be further discussed in the results section.)

The Transformer decoder layer involves three main sub-layers:
\begin{enumerate}
    \item \textit{Masked multi-head self-attention:} 
    Each future step \(\mathbf{q}_l \in \mathbf{Q}\) can attend to other steps \(\mathbf{q}_m\) within the query sequence, where \(m \le l\) in a causal (masked) manner. The goal is to prevent a given step from peeking at future positions beyond itself.
    \item \textit{Cross-attention with encoder output:} 
    After self-attention in the decoder, the resulting target representations \(\mathbf{Q}^{\prime}\) attend to the encoder output \(\mathbf{H}^{(N_{\mathrm{enc}})}\) to incorporate information about the historical sequence resulting $\mathbf{Q}^{\prime\prime}$.
    This sub-layer fuses future-step queries with historical context learned by the encoder.
    \item \textit{Feed-forward network:} 
    Finally, each token in \(\mathbf{Q}^{\prime\prime}\) is passed through a position-wise feed-forward network which typically comprises two linear layers and a nonlinear activation. Residual connections and layer normalization are applied around each of the three sub-layers (self-attention, cross-attention, and feed-forward network) to stabilize and speed up training.
\end{enumerate}
Similarly, positional encodings are incorporated into the target embeddings \(\mathbf{Q}\) to convey sequence order information.

Putting it all together, a decoder layer denoted by \(\mathtt{g}^{\text{dec},(k)}\) for layer $k$, transforms its input \(\mathbf{Q}^{(k-1)}\) into an output \(\mathbf{Q}^{(k)}\) via:
\begin{equation}
    \mathbf{Q}^{(k)} = \mathtt{g}^{\text{dec},(k)}\!\bigl(\mathbf{Q}^{(k-1)}, \mathbf{H}^{(N_{\mathrm{enc}})}\bigr),
    \quad
    k = 1, \ldots, N_{\mathrm{dec}},
\end{equation}
where \(N_{\mathrm{dec}}\) is the total number of decoder layers. The final decoder output \(\mathbf{Q}^{(N_{\mathrm{dec}})}\) can then be projected to the desired parameter space. Specifically, we employ another fully connected layer mapping each token in \(\mathbf{Q}^{(N_{\mathrm{dec}})}\) to \(\boldsymbol{\theta}_{n+l}\), where \(l\in\{0,\dots,L-1\}\). Hence, the decoder produces 
$\boldsymbol{\Theta}  \in\;\mathbb{R}^{L \times V} $.
where each row fully specifies the parameters of the delay distribution for the corresponding future packet.

The encoder-decoder Transformer effectively captures historical trends and future dependencies by decoupling past encoding from future decoding, enforcing causality, and enabling parallel computations. This design is especially suited for multi-step predictions of packet delay distributions, offering scalability.

In the following subsection, we examine the final probabilistic representation layer, which uses the raw parameters $\boldsymbol{\Theta}$ to compute the delay probabilities.

\subsection{Parametric Distribution Prediction}

\begin{figure}[ht]
    \centering
    \includegraphics[width=0.35\textwidth, trim={5.2cm 0.2cm 3.7cm 0.3cm}, clip]{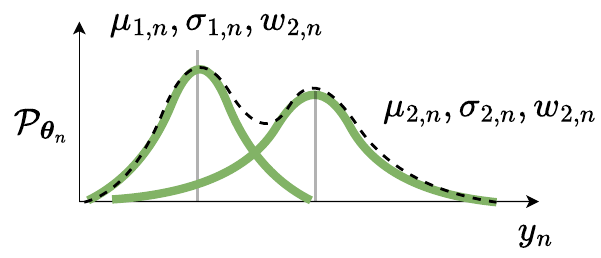}
    \caption{Gaussian mixture model example with 2 components and their parameters denoted by $\mu$ as mean, $\sigma$ as variance, and $w$ as mixture weight.}
    \label{fig:gmm}
\end{figure}

Rather than predicting a single point estimate of future delay, we aim to model the \emph{entire} delay distribution. Such distributional forecasts are vital for time-sensitive applications that require robust estimates of worst-case or high-percentile latencies. This task falls under the umbrella of conditional density estimation (\gls{CDE}), where we learn a mapping from input conditions (historical and contextual data) to the parameters of a chosen distribution.

In the present work, we explore \gls*{MDN} with \gls*{GMM} that can be used to capture potentially multimodal delay behaviors, characterized by a set of weights, means, and variances as shown via a \gls{GMM} with two centers in Figure \ref{fig:gmm}. Mathematically, we denote the parametric distribution for packet \(n+l\) as $\mathcal{P}_{\boldsymbol{\theta}_{n+l}}(y_{n+l})$, where \(\boldsymbol{\theta}_{n+l}\) are parameters output by the neural network. Once these parameters are obtained, we can derive quantities of interest (e.g., mean, variance, or percentile thresholds such as the 95\textsuperscript{th} percentile) to facilitate admission control, resource allocation, and other \gls{TSN}-driven optimization tasks.

Having outlined the complete trainable pipeline for probabilistic prediction, we now describe how the network is trained using supervised learning methods.

\subsection{Dataset Creation and Training} \label{sec:training_inference}

To train the proposed models, we first build a dataset of historical sequences and corresponding future delays. Each training sample contains:
\begin{itemize}
    \item \emph{Historical window}: A sequence of $H$ past delay context vectors \( \mathbf{X}_{m} = \{\mathbf{x}_{m-i+1}\}_{i=1}^{H}\), which encapsulates the conditions up to index $m$.
    \item \emph{Future window}: The subsequent $L$ delay values \( \{y_{m}, y_{m+1}, \dots, y_{m+L-1}\}\), each of which the model seeks to predict via a parametric distribution with parameters \(\boldsymbol{\theta}_{m+l}\) for $\forall l \in \{ 0:L-1 \}$.
\end{itemize}
We thus form a training set 
\begin{equation}
    \mathcal{D} 
    \;=\;
    \bigl\{ 
        (\mathbf{X}_m,\, y_{m},\, \dots,\, y_{m+L-1}) 
    \bigr\}_{m=1}^{N},
\end{equation}
where $N$ is the total number of samples. Note that each $\mathbf{X}_m$ is associated with $L$ future delay values, enabling multi-step supervision in a single training pass.

Given a parametric distribution 
\(\mathcal{P}_{\boldsymbol{\theta}_{m+l}}(\cdot)\)
for each future delay \(y_{m+l}\), the model outputs 
\(\boldsymbol{\theta}_{m+l} \in \mathbb{R}^V\)
for \(l = 1,\dots,L\). We minimize the negative log-likelihood (NLL) across all future steps:
\begin{equation}
    \label{eq:loss}
    \mathcal{L}(\mathcal{D}) 
    \;=\; 
    - \sum_{m=1}^{N} \sum_{l=0}^{L-1} 
    \ln \!\Bigl( \mathcal{P}_{\boldsymbol{\theta}_{m+l}} \bigl(y_{m+l}\bigr) \Bigr),
\end{equation}
where 
\(\boldsymbol{\theta}_{m+l} = [g_{\mathbf{W}}\bigl(\mathbf{X}_m\bigr)]_{l}\)
incorporates the tokenization and temporal modeling (via LSTM or Transformer). Here, $g_{\mathbf{W}}(\cdot)$ is the trainable mapping that produces the distribution parameters for each future time step. By minimizing (\ref{eq:loss}), the model is encouraged to match the full distribution of the delay, rather than just a single statistic.


By simultaneously fitting the parameters of a parametric distribution and leveraging temporal context, the model learns both short- and long-range dependencies in packet delay. As conditions evolve in online scenarios, newly observed samples can be integrated to continually refine model estimates, ensuring robust and adaptive latency prediction in 5G networks.

\subsection{Summary of the Proposed Approach}

In this work, we emphasize three key innovations compared to standard neural 
network predictors introduced in the established machine learning approaches 
that inspired our work:

\begin{enumerate}
    \item \textbf{Learnable tokenization of heterogeneous features.}
    For each time step, we embed both continuous and discrete network 
    indicators (e.g., packet size, MCS index, or HARQ counts) into a 
    single, learnable token. This avoids manual feature engineering 
    while capturing the critical context for delay prediction.
    
    \item \textbf{Full probabilistic forecasting.}
    Instead of predicting a single point estimate for the delay, 
    we attach a mixture density layer that outputs parameters of 
    a distribution. This allows us to capture the entire delay profile, 
    including tail latencies and other rare, high-delay events.
    
    \item \textbf{Parallel generation for faster inference.}
    In both LSTM and Transformer backbones, we adopt a parallel 
    “decoding” procedure. For the LSTM, we feed padding tokens 
    during generation steps; for the Transformer, we use a masked 
    parallel decoder. In both cases, training time is reduced compared to autoregressive decoding, without compromising accuracy.
\end{enumerate}

During online deployment, the model continually updates its input sequence with newly observed packet delays and network conditions, refining its understanding of the system as it evolves. This adaptive nature is particularly relevant in 5G networks, where user mobility and varying traffic loads can alter channel conditions and resource availability over time.

The following sections detail how this methodology is implemented and validate its effectiveness via real 5G measurements and performance evaluations.

\section{Evaluation Methodology}
\label{sec:evaluation}

In this section, we detail the experimental setup used to assess our proposed approaches and explain the evaluation methodology. This includes the implementation of the proposed approaches, the key performance metrics for delay prediction, the model variants, and their implementation. Additionally, we describe the different configurations and comparison schemes used in our experiments.

\subsection{Data Collection Setup}

\begin{figure}
	\centering
	\begin{subfigure}{0.32\textwidth}
		\includegraphics[width=\linewidth]{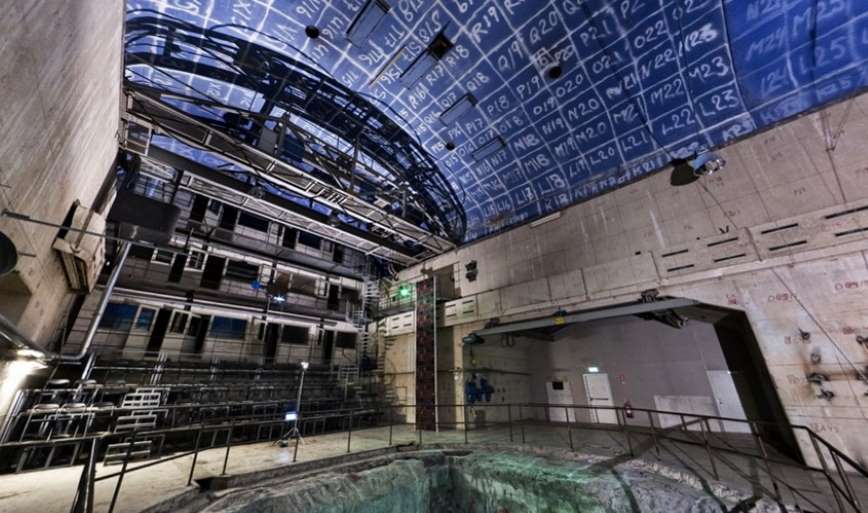}
            \label{fig:calib5}
	\end{subfigure}
	\begin{subfigure}{0.15\textwidth}
		\includegraphics[width=\linewidth]{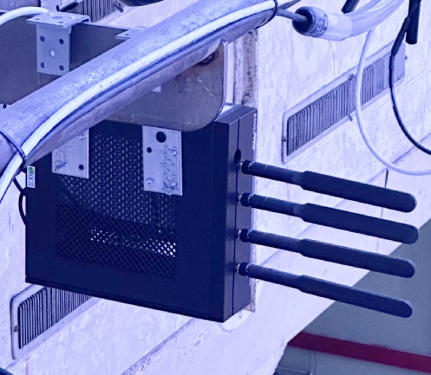}
            \label{fig:calib20}
	\end{subfigure}
	\caption{Location of the experiments at the ExPECA testbed in the R1 hall at KTH University (left), with an SDR node used for measurements (right).}
	\label{fig:r1}
\end{figure}
All proposed machine learning models in this work (LSTM, Transformer, MLP, etc.) were implemented in \texttt{PyTorch} \footnote{Python code and the datasets could be found at: https://github.com/samiemostafavi/wireless-tpp}. We used a uniform training and inference pipeline, ensuring that any variations in model performance arise from architectural differences rather than implementation details.
Training and evaluations were conducted on two Dell R450 servers, each equipped with an NVIDIA L4 GPU featuring 24GB memory, an Intel(R) Xeon(R) Silver 4314 CPU operating at 2.40GHz, and 128GB RAM.

To collect real-world data, we employed an SDR-based 5G system using \emph{OpenAirInterface5G} on the ExPECA testbed \cite{oai2020, mostafaviExPECA}. 
ExPECA is designed as an edge-computing test environment, comprising servers and SDRs connected through a flexible networking fabric, all synchronized in time using the Precision Time Protocol (PTP).
These components are deployed in an underground facility with no external interference, making it well-suited for reproducible, end-to-end, time-sensitive wireless experiments.
Figure \ref{fig:r1} shows the location of ExPECA testbed and an SDR node used for measurements.
We elaborate about the capabilities and implementation details of ExPECA in our previous work in \cite{mostafaviExPECA}.

We also leveraged the previously published work \emph{EDAF} (End-to-End Delay Analytics Framework) to gather and preprocess the end-to-end delay measurements and associated context (e.g., packet size, MCS index, number of retransmissions) \cite{mostafaviedaf}.
The \emph{EDAF} framework can extract comprehensive timely information about the 5G processes that manage packet transmission, thus facilitating an in-depth analysis of network performance with respect to end-to-end (E2E) delay.
Its core objective is to capture timestamps and key metadata from the packet’s journey in realtime, starting at the application node (UE side) and ending at the server node (gNB side).
The data collection setup is similar to Figure \ref{fig:systemmodel}, and our experiments were conducted with the following specifications.
\begin{itemize}
    \item \textbf{Deployment Setup:} 
    We ran experiments on two high level nodes within the ExPECA testbed: \texttt{SDR-04} (gNB) and \texttt{SDR-07} (UE). This link has line of sight connection with 20m distance and 10m above the ground.
    \item \textbf{Connection Quality:} 
    Two radio gain configurations were used: one maintained a stable uplink MCS index fixed at 20, achieving an RSRP of -90\,dBm with minimal HARQ retransmissions (stable high-gain configuration), while the other reduced the gains, resulting in an MCS index that fluctuated between 12 and 18 and a BLER of approximately 10\% (reduced gain configuration).
    \item \textbf{Traffic Patterns:}
    In stable high-gain configuration measurements, we had constant 100B packets and inter-arrival times of 10ms, 20ms, 50ms and 100ms.
    In the reduced gain configuration measurements, we had constant 200B packets and 50ms interarrival times.
    \item \textbf{Data Volume:}
    Over the course of the experiments, we collected more than 1\,M packet samples along the same link, yielding a large and diverse dataset of delay measurements under different load conditions and channel states.
\end{itemize}

The final dataset was preprocessed (e.g., outlier filtering, synchronization, normalization of features) and stored for offline training and evaluation. 
We partitioned the data into training (80\%), validation (20\%), and test sets (separate larger set with the size of $50k$ samples) to facilitate hyperparameter tuning and unbiased performance assessments.

\subsection{Implementation Details and Model Variants}
\label{sec:implementation}

We evaluate several neural architectures for modeling end-to-end delay distributions, all sharing a common tokenization scheme and an \gls{MDN} output layer. In our approach, each packet's delay context vector includes the following features:
\begin{itemize}
    \item Packet size (continuous)
    \item Inter-arrival time (continuous)
    \item Packet arrival slot (discrete)
    \item \gls{MCS} index (discrete)
    \item \gls{HARQ} retransmission count (discrete)
    \item \gls{RLC} retransmission count (discrete)
\end{itemize}
Each feature is embedded into a uniform \(S\)-dimensional token (with \(S=16\) by default). Continuous features are normalized and projected through learnable linear layers, while discrete features are processed using embedding layers that handle missing or invalid entries via dedicated padding indices. A compact three-layer \gls{MLP} then expands, compresses, and projects the concatenated embeddings back to an \(S\)-dimensional vector, employing \texttt{Dropout} and \texttt{ReLU} activations to enhance generalization. Once the core neural network has processed the tokenized input, we append an \gls{MDN} head that estimates a Gaussian mixture distribution with 8 components (yielding \(V=24\) parameters). An affine transformation is applied to the delay dimension, and small Gaussian noise (standard deviation $0.1$) is injected during training to further improve robustness.


In designing our comparison schemes, we differentiate between two model categories—single-step and multi-step—based on our problem formulation. Multi-step models generate distinct parametric distributions for each future time step (e.g., the LSTM and Transformer models in our approach). In contrast, single-step models output a single set of parameters representing the distribution for all \(L\) future delays, as they lack the autoregressive structure required for multi-step forecasting. For single-step models, we modify the loss function (see Equation \ref{eq:loss}) by using \(\mathcal{P}_{\boldsymbol{\theta}_{m}}\) instead of \(\mathcal{P}_{\boldsymbol{\theta}_{m+l}}\), effectively treating one historical sequence as corresponding to a single parametric distribution. This categorization reflects prior work focused on single-step prediction, while our architecture is designed to efficiently address multi-step delay prediction.

\begin{table}[ht]
    \centering
    \caption{Models Used in Comparison Schemes}
    \label{tab:model-comparison}
    \begin{tabular}{|l|l|l|l|}
    \hline
    \textbf{Model Name} & \textbf{Processing History} & \textbf{Prediction} & \textbf{Parameters} \\
    \hline
    \textsc{MLP} & Single-step & Single-step & 37k \\
    \hline
    \textsc{LSTM-SS} & Multi-step  & Single-step & 33k \\
    \hline
    \textsc{LSTM} & Multi-step & Multi-step & 33k \\
    \hline
    \textsc{Transformer} & Multi-step & Multi-step & 78k \\
    \hline
    \end{tabular}
\end{table}

Based on this categorization, Table~\ref{tab:model-comparison} summarizes 
the core models and their approaches to processing historical data and 
predicting future delays. We consider four model variants: 
\begin{itemize}
    \item Single-step \gls{MLP} (MLP),
    \item Single-step \gls{LSTM} (LSTM-SS),
    \item Multi-step \gls{LSTM} (LSTM),
    \item Multi-step Transformer (Transformer).
\end{itemize}
All are configured with a hidden dimension of \(S = 16\) and a dropout rate of 0.2.

Our Transformer model adopts a standard encoder-decoder structure, featuring multi-head self-attention with 4 heads per layer, feed-forward sub-layers with a 32\(\times\) expansion, GELU activations, dropout, and residual connections. The default configuration uses 6 encoder layers and 6 decoder layers. For both single-step and multi-step \glspl{LSTM}, we follow conventional implementations by stacking 2 layers, as additional layers were found to degrade performance.

The single-step \gls{MLP} model is a fully connected feed-forward network that relies solely on the most recent context token. It consists of 4 fully connected layers with residual connections, dropout, and GELU activations, deliberately omitting longer historical context to serve as a lightweight baseline. In the single-step \gls{LSTM}, we extract the most recent hidden state \(\mathbf{h}_n\) from the \gls{LSTM} processing the history sequence and use a linear layer to transform it into the parameter vector.
These two models represent the state-of-the-art methods discussed in related work \cite{mostafaviDataDrivenLatencyProbability2023c,skocajDatadrivenPredictiveLatency2023a}. However, we have enhanced them by incorporating dropout, residual connections, and the tokenization layer to ensure a fair comparison and robust benchmarking.

A summary of the trainable parameter counts for our model variants is presented in Table \ref{tab:model-comparison} as well. The Transformer model, with its complex encoder-decoder architecture and multi-head self-attention, requires 78k parameters, which is higher than the 33k parameters used by both the single-step and standard \gls{LSTM} models, and the 37k parameters used by the \gls{MLP}. An important characteristic of these temporal neural architectures is that their parameter count remains almost fixed, regardless of the length of the historical or target sequences. This ensures that the model complexity does not increase with longer sequences, enhancing scalability.

Ultimately, we consider two additional aspects in our comparison schemes. First, we vary the lengths of the historical and future windows by training models on datasets with different (H : L) configurations—specifically, (10:10), (20:20), (50:50), and (100:100). This allows us to assess how prediction efficiency and performance change as the complexity of the temporal context increases. Second, we experiment with varying training set sizes—using 1k, 2.5k, 5k, and 10k samples—to examine the impact of data volume on model accuracy and determine the amount of data needed to achieve robust performance.

\subsection{Performance Metrics}

To evaluate each model’s ability to forecast future latency distributions under realistic 5G conditions, we use the following core KPIs:
\begin{itemize}
    \item \textbf{Negative Log-Likelihood (NLL):}
    This standardized measure for probabilistic modeling also serves as our training loss. Lower NLL values indicate that the predicted distributions more closely align with the observed delay data.

    \item \textbf{Mean Absolute Error (MAE):}
    While the focus is on distribution-based prediction, MAE still provide insight into average prediction error compared to ground truth.

    \item \textbf{Coverage Probability (e.g., 99\% Coverage):}
    Captures how frequently the actual delay falls within the model’s high-confidence region (e.g., the 99th percentile) of its predicted distribution.
\end{itemize}

Beyond accuracy, we evaluate efficiency metrics crucial for practical deployment. We track \textbf{training requirements}, including the amount of data needed to reach stable performance and the computational overhead of training. \textbf{Model size} is measured by counting parameters and analyzing its impact on performance and training time as the historical window size \(H\) and prediction horizon \(L\) increase. Models that require excessive capacity or slow down significantly with larger \(H\) or \(L\) may be impractical in real-world applications.

\section{Numerical Results} \label{sec:results}

\begin{figure}
	\centering
	\begin{subfigure}{0.48\textwidth}
		\includegraphics[width=\linewidth]{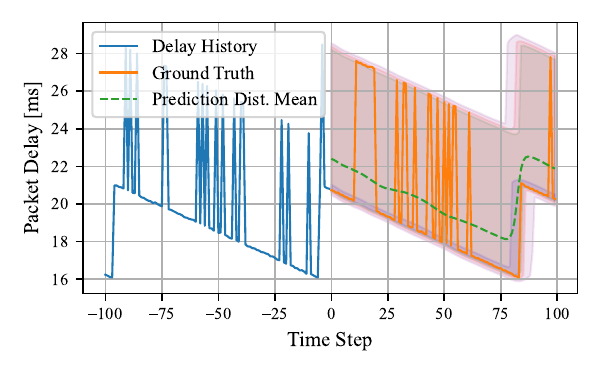}
		\caption{Multi-step transformer predictor}
            \label{fig:proofa}
	\end{subfigure}
	\begin{subfigure}{0.48\textwidth}
		\includegraphics[width=\linewidth]{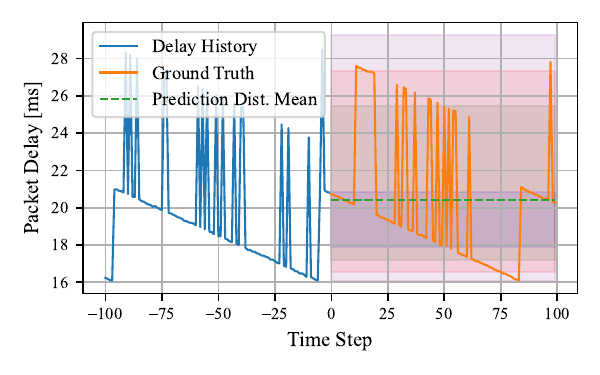}
		\caption{Single-step \gls{MLP} predictor}
            \label{fig:proofb}
	\end{subfigure}
	\caption{Comparison of multi-step transformer and single-step \gls{MLP} predictors on a test sample with 100 historical and 100 future time steps, trained on a 10k dataset with reduced gain configuration. The multi-step model produces time-varying distributions (colored regions for 50\%, 70\%, 90\%, and 99\% coverage, from dark to light) that closely cover the ground truth, while the single-step model yields uniform forecasts.}
	\label{fig:proof}
\end{figure}

We begin our evaluations by examining experiments conducted under the reduced gain configuration. In these experiments, retransmissions occur more frequently and link adaptation results in an average MCS index of 15 (ranging from 12 to 18). First, we present a proof-of-concept plot to motivate the use of multi-step probabilistic delay prediction, which is the central focus of this work. Figure \ref{fig:proof} shows a prediction sample from the test dataset comprising 200 time steps—100 steps of historical data (representing 5 seconds) and 100 steps forming the prediction horizon (covering the next 5 seconds). Although the figure displays only the past delay values, the predictor actually receives a sequence of 100 packet delay context vectors as input.

The delay time series in Figure \ref{fig:proof} reveals two notable effects. First, the delay follows a sawtooth pattern, which we attribute to frame-alignment delays. In our setup, packets are generated every 50 ms, but because the application clock is not synchronized with the 5G clock source, there is a varying shift in packet arrival times relative to the frame start. This misalignment modulates the scheduling handshake delays associated with TDD patterns (for further details, see \cite{mostafaviedaf}). Second, the series exhibits random jumps of approximately 7.5 ms, corresponding to retransmissions that increase the overall packet delay.

On the right side of both sub-figures in Figure \ref{fig:proof}, the probabilistic predictions are depicted as colored regions representing coverage intervals at 50\%, 70\%, 90\%, and 99\% confidence levels (from darker to lighter shades). These models were trained using the same 10k training dataset. A comparison between the multi-step and single-step predictors demonstrates the advantage of multi-step forecasting. The transformer-based multi-step model closely tracks the temporal dynamics of the ground truth, while the single-step \gls{MLP} predictor tends to treat all future time steps similarly. For example, as shown in Figure \ref{fig:proofa}, the multi-step model indicates that, at step 75, the likelihood of the delay exceeding 24 ms is very low—an insight that the single-step predictor fails to capture. This example illustrates how multi-step probabilistic predictions can enable more proactive adaptation in dynamic network environments.

\begin{figure}[ht]
    \centering
    \includegraphics[width=0.48\textwidth]{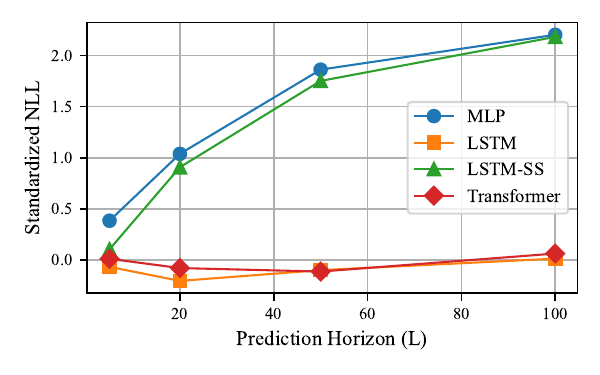}
    \caption{Comparison of model accuracy across different prediction horizons. All models were trained on a reduced gain configuration dataset containing 5k samples.}
    \label{fig:proofnll}
\end{figure}

In Figure~\ref{fig:proofnll}, we compare the standardized negative log-likelihood (NLL) for four models (MLP, LSTM, LSTM-SS, and Transformer) as the prediction horizon \(L\) extends from 10 to 100. All models were trained on a 5k-sample dataset. At shorter horizons, the difference among the models is minimal, indicating that single-step models can be sufficient when only a few future time steps need to be predicted. However, as \(L\) grows, the single-step variants (MLP and LSTM-SS) exhibit a pronounced rise in NLL, suggesting they struggle to preserve accurate probabilistic forecasts over longer windows. By contrast, the multi-step models (LSTM and Transformer) maintain a lower NLL across all horizons, reflecting their stronger ability to capture temporal dependencies.

Furthermore, in Figure~\ref{fig:proofnll}, LSTM-SS outperforms the MLP by effectively capturing historical trends that the MLP fails to incorporate. We also observe a modest increase in NLL for the LSTM and Transformer models as the prediction horizon \(L\) grows. This trend, which will be further examined in later benchmarks, reflects the natural decline in predictability as forecasts extend further into the future, thereby diminishing the influence of historical observations.
We also evaluated the MAE for the four models. The single-step models show a slight upward trend in MAE—from 2.61\,ms to 2.68\,ms—as \(L\) increases, while the multi-step models consistently achieve an MAE of about 2.6\,ms across all horizons. These differences are minimal, likely due to the broad delay distribution and the fact that the predicted expected values, whether from single-step or multi-step approaches, are noticeably offset from the ground truth (see Figure~\ref{fig:proofnll}).

\begin{figure}[ht]
    \centering
    \includegraphics[width=0.48\textwidth]{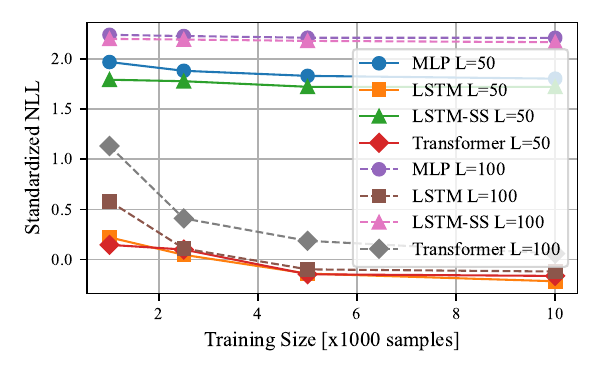}
    \caption{Comparison of model accuracy across training dataset size. All models were trained on a reduced gain configuration dataset.}
    \label{fig:proofdatasetsize}
\end{figure}

Next, we investigate how varying the training dataset size impacts the test NLL of our models. Figure~\ref{fig:proofdatasetsize} shows the test NLL for different training dataset sizes using the same experimental configuration as before. We observe that models forecasting further into the future ($L=100$) exhibit higher NLL values, which is expected due to increased uncertainty. The single-step models perform worse overall, although their performance improves marginally with more training data. In contrast, the multi-step models benefit more substantially from additional samples—showing a significant improvement when increasing the dataset from 1k to 5k samples, with further gains from 5k to 10k being less pronounced. Moreover, models with a longer prediction horizon (\(L=100\)) require more data to achieve stable performance compared to those with a shorter horizon (e.g., \(L=50\)).

In the next phase of our evaluations, we examine data collected from experiments conducted under a stable high-gain configuration. In this setting, the delay profile is less stochastic because retransmissions are rare. However, we varied the packet interarrival times in each experiment to build a comprehensive dataset that enables our models to capture the impact of traffic periodicity—a key feature in the delay context vector. In this dataset, packets are 200B in size and generated at periodicities of 10, 15, 20, 25, 50, and 100 ms (equally contributing to the dataset). Since most periodicities are faster than 50 ms and the packet size is doubled, we expect more pronounced temporal dependencies and queuing effects in this configuration.

\begin{figure}[ht]
	\centering
	\begin{subfigure}{0.48\textwidth}
		\includegraphics[width=\linewidth]{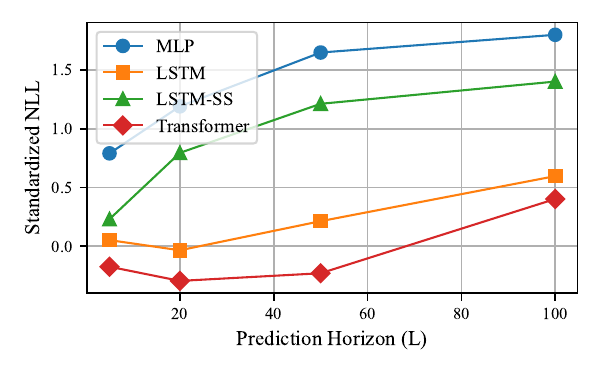}
		\caption{Accuracy in terms of \gls{NLL}}
            \label{fig:nllintervals}
	\end{subfigure}
        \begin{subfigure}{0.48\textwidth}
		\includegraphics[width=\linewidth]{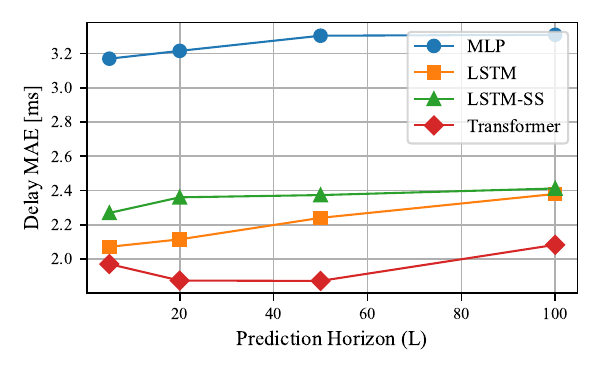}
		\caption{Accuracy in terms of \gls{MAE}}
            \label{fig:maeintervals}
	\end{subfigure}
	\caption{Comparison of model accuracy across different prediction horizons. All models were trained on a stable high gain configuration dataset containing 5k samples.}
	\label{fig:accintervals}
\end{figure}

Figures~\ref{fig:nllintervals} and~\ref{fig:maeintervals} summarize the prediction performance of our models under the stable high-gain configuration. Across all prediction horizons, the Transformer achieves the lowest standardized negative log-likelihood (NLL), demonstrating its superior ability to capture temporal dependencies and the inherent variability in the delay process. This advantage is particularly evident in experiments with rapid packet transmissions, where long-term dependencies are more pronounced. The multi-step LSTM ranks second by leveraging its autoregressive structure to follow sequential patterns, although its NLL noticeably increases from \(L=50\) to \(L=100\), reflecting the growing challenge of longer-term prediction. In contrast, the single-step models (LSTM-SS and MLP) underperform relative to the multi-step variants; however, LSTM-SS shows a marked improvement over previous reduced-gain experiments by effectively utilizing historical data. Figure~\ref{fig:maeintervals} depicts the mean absolute error (MAE), where the Transformer again consistently outperforms the other models, and the LSTM maintains an intermediate level of accuracy. As the prediction horizon \(L\) increases, the feedforward MLP struggles the most, highlighting the benefits of recurrent and attention-based architectures in modeling temporal dependencies.

\begin{figure}[ht]
    \centering
    \includegraphics[width=0.48\textwidth]{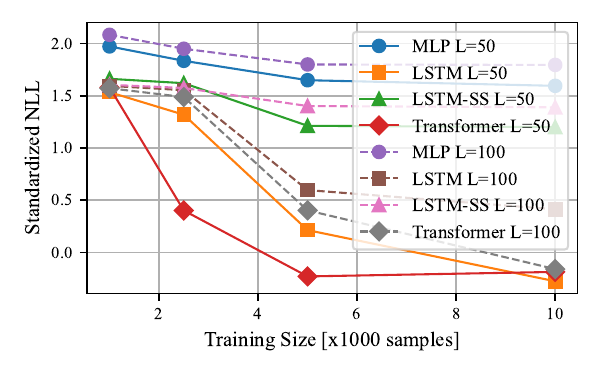}
    \caption{Comparison of model accuracy across training dataset size. All models were trained on a stable high gain configuration dataset.}
    \label{fig:intervalsdatasetsize}
\end{figure}

Next, we assess the impact of varying training dataset sizes on the test NLL of our models under the high gain configuration. Figure~\ref{fig:intervalsdatasetsize} displays the test NLL for different training dataset sizes. Similar to Figure~\ref{fig:proofdatasetsize}, models forecasting further into the future (\(L=100\)) exhibit higher NLL values due to increased uncertainty. However, in this setting, performance improvement saturates later, with most models achieving optimal performance only when trained on a 10k dataset. The single-step models still perform worse overall, though they benefit more noticeably from additional training data compared to Figure~\ref{fig:proofdatasetsize}. Likewise, the multi-step models show significant improvements when the dataset size increases from 1k to 5k and from 5k to 10k.

\begin{figure}[ht]
    \centering
    \includegraphics[width=0.48\textwidth]{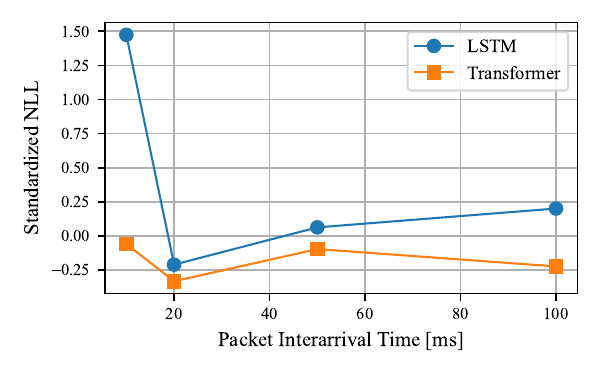}
    \caption{Decomposed test \gls{NLL} conditioned on various packet inter-arrival times. Models were trained over a datasets of size 5k from stable high gain configuration.}
    \label{fig:ia}
\end{figure}

In Figure~\ref{fig:ia}, we present the standardized \gls{NLL} for test samples grouped by similar packet interarrival times. This analysis helps us determine whether model inaccuracies are tied to specific conditions—in this case, varying interarrival times—or if some sample categories are inherently more challenging to predict. The figure shows results for models trained on a 5k-sample dataset under a stable high gain configuration. Notably, the LSTM model struggles to predict delays accurately when packets arrive rapidly (e.g., at 10ms interarrival), highlighting the increased temporal coupling in such scenarios. By contrast, the Transformer consistently delivers lower \gls{NLL} values, particularly under challenging conditions, underscoring the robustness of attention-based architectures in capturing complex dependencies even in high-intensity traffic.

\begin{figure}
	\centering
	\begin{subfigure}{0.48\textwidth}
		\includegraphics[width=\linewidth]{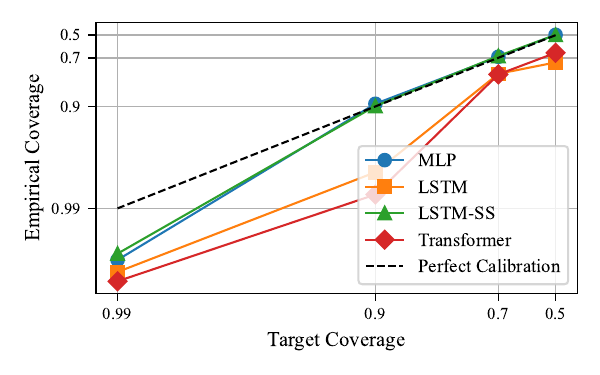}
		\caption{Reduced gain configuration}
            \label{fig:calibred}
	\end{subfigure}
	\begin{subfigure}{0.48\textwidth}
		\includegraphics[width=\linewidth]{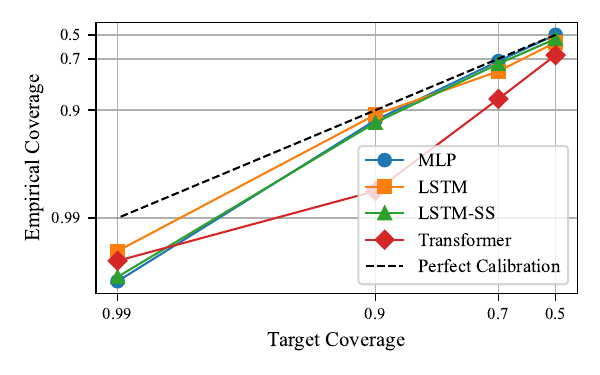}
		\caption{Stable high gain configuration}
            \label{fig:calibhigh}
	\end{subfigure}
	\caption{Empirical coverage derived using different predictors trained on 10k datasets and $L=50$.}
	\label{fig:calib}
\end{figure}

Next, we assess the empirical coverage of the predicted delay distributions. Empirical coverage measures how often the actual delay falls within a specific confidence interval generated by the model. For example, for a 50\% nominal coverage, we compute the 25th and 75th percentiles of the predicted distribution, and ideally, 50\% of the ground truth delay values should lie outside this interval if the model is perfectly calibrated. Figure~\ref{fig:calib} illustrates the empirical coverage of various models at a prediction horizon \(L=50\) using a 10k-sample dataset. The dashed diagonal line represents perfect calibration, where the empirical coverage matches the nominal target \((1-\alpha)\). Deviations below the line indicate over-coverage (the predicted intervals are too wide), while deviations above the line indicate under-coverage (the predicted intervals are too narrow).

Notably, although the single-step models tend to underperform in terms of NLL, they achieve closer-to-ideal calibration at lower confidence levels. In contrast, the better-performing models—such as the LSTM and Transformer in the reduced gain experiments, and the Transformer in the high gain setup—tend to over-cover across all confidence levels by producing wider prediction intervals. While over-coverage is generally less risky than under-coverage, it may lead to inefficiencies in certain applications. This pattern holds across different training dataset sizes, with models that achieve lower NLL also exhibiting higher levels of over-coverage. However, for higher confidence levels (e.g., 99\%), all models perform similarly, though in the more challenging high gain scenario, the LSTM and Transformer models come closer to perfect calibration.

\begin{figure}
    \centering
    \includegraphics[width=0.48\textwidth]{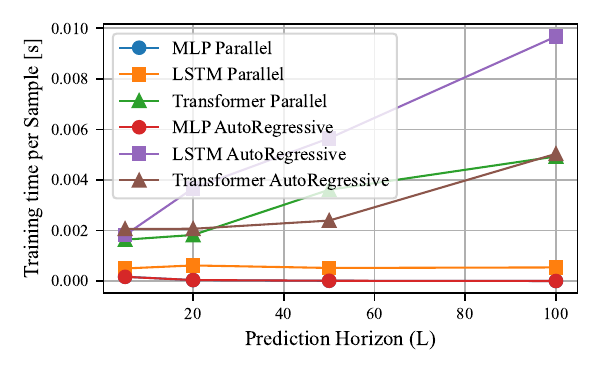}
    \caption{Per-sample training time (in seconds) for MLP, LSTM, and Transformer models in both parallel and autoregressive configurations.}
    \label{fig:trainingtime}
\end{figure}

Figure \ref{fig:trainingtime} presents the per-sample training times (in seconds) for MLP, LSTM, and Transformer architectures evaluated under both parallel and autoregressive setups as the prediction horizon (L) varies from 5 to 100, aiming to assess the computational complexity of these models. The MLP consistently incurs the lowest cost due to its simple architecture and direct output generation, while the parallel LSTM maintains relative efficiency through the partial parallelization we applied. In contrast, the training time for the autoregressive LSTM increases dramatically with longer horizons due to its scalability limitations. Although Transformer variants exhibit higher per-sample training times overall, they still train faster than the autoregressive LSTM, largely because the computational expense of multi-head self-attention is better managed. Interestingly, the parallel Transformer—expected to process the entire target sequence in one pass and therefore significantly outperform its autoregressive version—shows only a modest improvement, a discrepancy we attribute to the internal causal masking described in our approach. As the prediction horizon expands, these trends become more pronounced, with both the autoregressive LSTM and Transformer variants displaying substantial increases in training time, highlighting the combined impact of model complexity and decoding strategy on overall computational cost.

\begin{figure}[ht]
    \centering
    \includegraphics[width=0.48\textwidth]{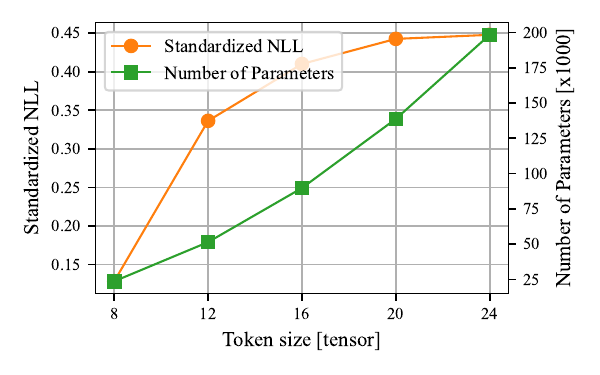}
    \caption{Model accuracy and size for the transformer model trained with different token sizes. Number of training samples are $50$k on the high gain configuration with $L = 50$.}
    \label{fig:token}
\end{figure}

Finally, in Figure~\ref{fig:token}, we examine the impact of token size (embedding dimension) on the accuracy and parameter efficiency. We test this on the transformer model. The results reveal a clear trade-off: increasing the token size reduces the standardized \gls{NLL}, improving the model’s ability to capture complex patterns and dependencies. However, this improvement comes at the cost of a quadratic increase in the number of model parameters, leading to larger memory requirements and potential overfitting risks. For example, moving from a token size of 8 to 24 reduces the \gls{NLL} significantly, but also increases the number of parameters from approximately 25\,k to 200\,k. This trade-off highlights the importance of selecting an appropriate token size that balances performance and computational efficiency. For the dataset used in this study, a token size of 16 achieves a favorable balance between performance and model complexity.

\section{Conclusions} \label{sec:conclusion}

In this work, we presented a novel probabilistic latency prediction framework for 5G networks that leverages deep temporal neural architectures, including LSTM and Transformer models, to forecast end-to-end delay distributions over multiple future time steps. By employing a tokenization scheme to encode rich historical network context, our approach was evaluated on a real-world SDR-based 5G testbeds. The experimental results demonstrate that multi-step predictors, particularly the Transformer, consistently achieve lower negative log-likelihood and mean absolute error values compared to single-step models, effectively capturing the evolving nature of wireless delays.

Our findings indicate that the advantages of multi-step forecasting become increasingly pronounced as the prediction horizon extends. While single-step models may suffice for short-term predictions, they struggle with longer horizons due to their limited capacity to adapt to changing network conditions. In contrast, multi-step models such as the LSTM and Transformer not only track temporal dynamics more accurately but also provide more reliable uncertainty estimates. This is crucial for enabling proactive network management strategies such as adaptive scheduling and resource allocation in time-sensitive environments.

Furthermore, our analysis of training data requirements reveals that although all models benefit from larger datasets, multi-step architectures derive substantial improvements in predictive performance without increasing model complexity. This fixed parameter scalability makes our framework particularly attractive for practical deployment in dynamic 5G and future 6G networks. Future work will focus on refining these architectures further and extending the evaluation to additional real scenarios, ultimately contributing to more robust quality-of-service guarantees.

\bibliographystyle{ieeetr}
\bibliography{refs.bib}

\begin{IEEEbiography}[{\includegraphics[width=1in,height=1.25in,clip]{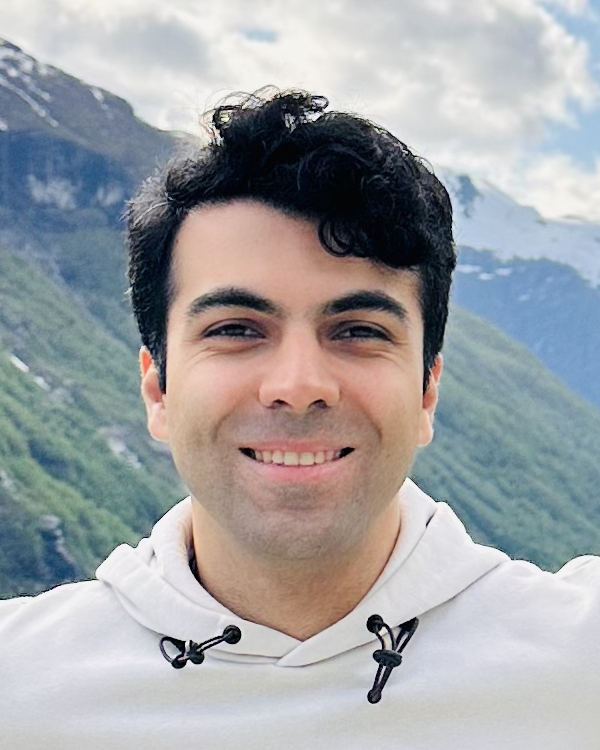}}]
{Samie Mostafavi} (Member, IEEE) received the bachelor’s degree in electrical engineering from the University of Tehran, in 2015, and the master’s degree in communication systems from KTH Royal Institute of Techonology in 2019.
He is currently pursuing the Ph.D. degree with the Department of Intelligent Systems, School of Electrical Engineering and Computer Science (EECS), KTH Royal Institute of Technology. 
He was with Ericsson AB Radio Research for a year before his doctoral studies which led to his M.Sc. thesis about vehicular positioning using 5G.
His current research is in the area of mobile edge computing and performance characterization, with a specific focus on data-driven approaches to predict delay in wireless communication systems.
\end{IEEEbiography}

\begin{IEEEbiography}[{\includegraphics[width=1in,height=1.25in,clip,trim={0cm 0 0cm 0}]{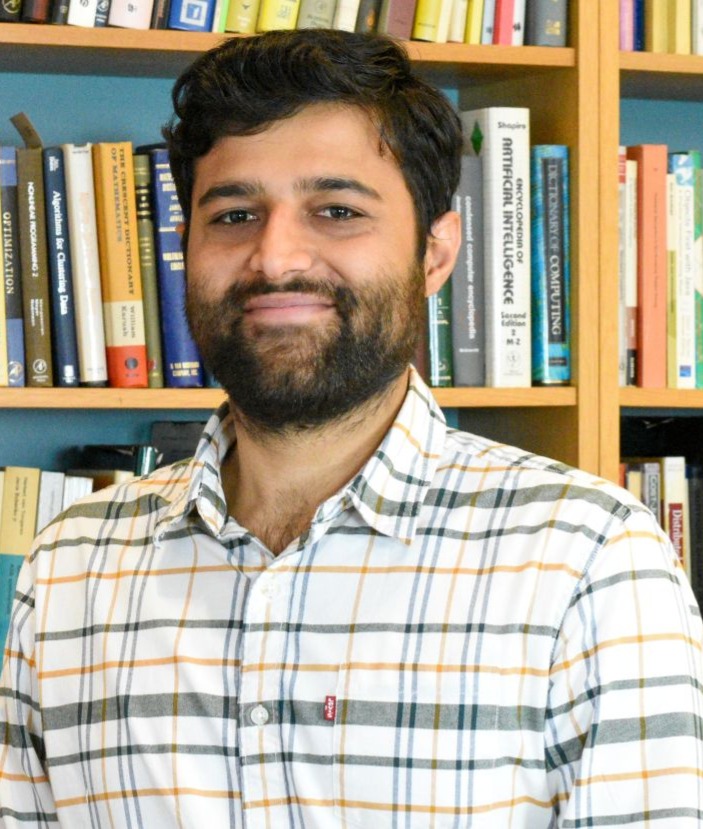}}]
{Gourav Prateek Sharma} has been an assistant professor at Thapar Institute of Engineering and Technology in Patiala, India, since December 2024. Prior to this appointment, he worked as a postdoctoral researcher in the ISE Division at the School of Electrical Engineering and Computer Science, KTH Royal Institute of Technology in Sweden. Dr. Sharma received his Ph.D. from IDLab at Ghent University in June 2022, where his research centered on developing and analyzing optimization algorithms for efficient resource allocation in telecommunications and media broadcast networks. He also holds a Master’s degree in Optoelectronics and Optical Communications from IIT Delhi (2017) and a Bachelor’s degree in Electronics and Communication Engineering from NIT Srinagar (2015).
\end{IEEEbiography}

\begin{IEEEbiography}[{\includegraphics[width=1in,height=1.2in,clip,trim={0cm 2cm 0cm 2cm}]{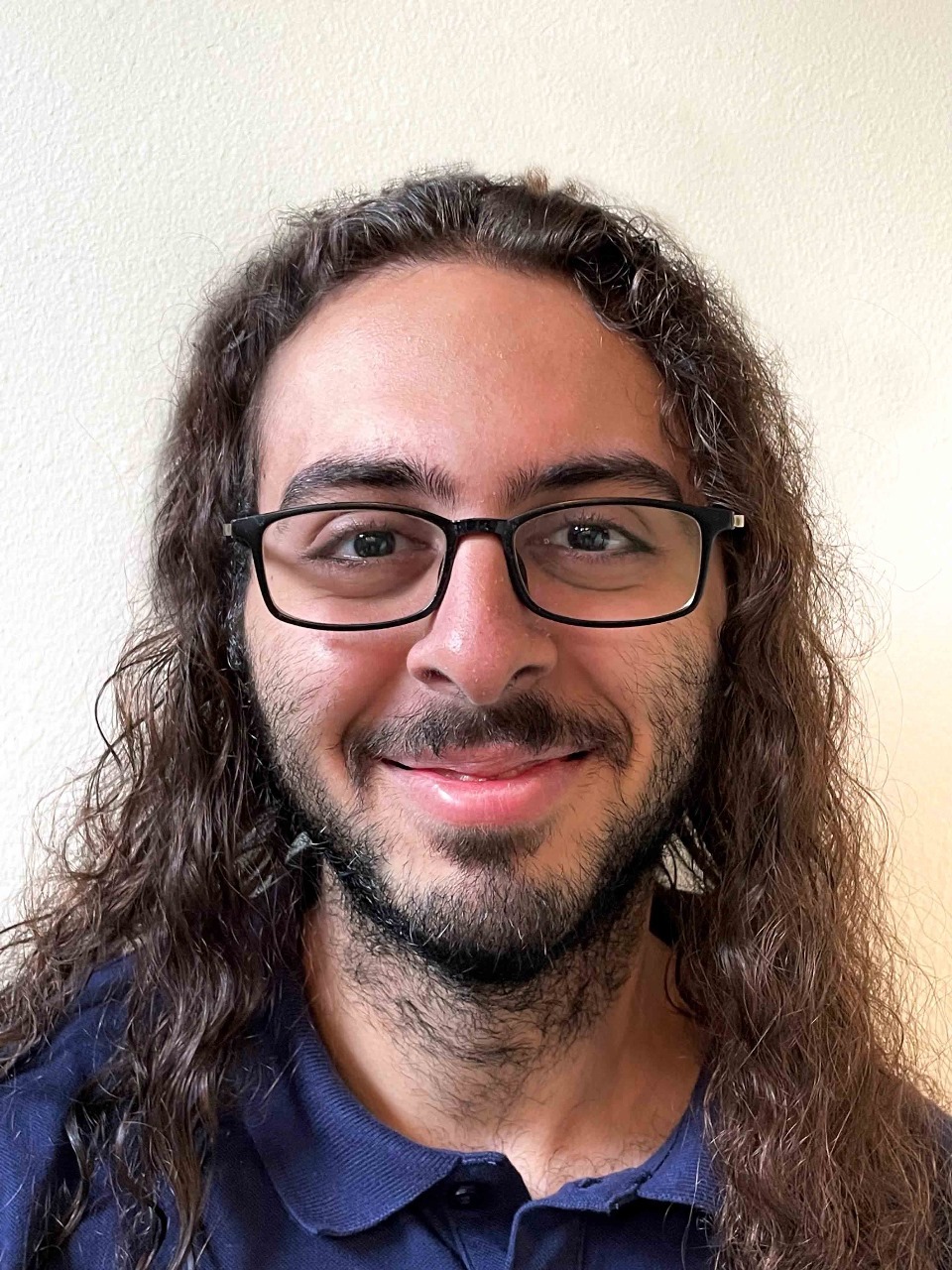}}]
{Ahmad Traboulsi} received a bachelor's degree in computer science from the American University of Beirut, in 2021, and is currently pursuing his master's degree in embedded systems at KTH Royal Institute of Technology. Along his studies, he is currently working as a research assistant at KTH, division of information science and engineering (ISE).
\end{IEEEbiography}

\begin{IEEEbiography}[{\includegraphics[width=1in,height=1.25in,clip]{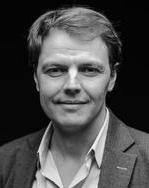}}]
{James Gross} (Senior Member, IEEE) received the Ph.D. degree from TU Berlin in 2006. 
From 2008 to 2012, he was with RWTH Aachen University, as an Assistant Professor and a Research Associate with the RWTH’s Center of Excellence on Ultra-High Speed Mobile Information and Communication (UMIC). 
Since November 2012, he has been with the Electrical Engineering and Computer Science School, KTH
Royal Institute of Technology, Stockholm, where he is a Professor of machine-to-machine communications. 
At KTH, he was the Director of the ACCESS Linnaeus Centre, from 2016 to 2019, while he is currently the Associate Director of the newly formed KTH Digital Futures Research Center, and the Co-Director of the newly formed VINNOVA Competence Center on Trustworthy Edge Computing Systems and Applications (TECoSA). 
He has authored over 150 (peer-reviewed) papers in international journals and conferences. 
His research interests are in the area of mobile systems and networks, with a focus on critical machine-to-machine communications, edge computing, resource allocation, and performance
evaluation. 
His work has been awarded multiple times, including the Best Paper Awards at ACM MSWiM 2015, IEEE WoWMoM 2009, and European Wireless 2009. 
In 2007, he was a recipient of the ITG/KuVS Dissertation Award for the Ph.D. Thesis.
\end{IEEEbiography}

\end{document}